\documentstyle[epsf,prl,aps,floats]{revtex}

\begin{document}
\draft

\wideabs{

\title{Reentrant charge order transition in the extended Hubbard model}

\author{R.\ Pietig, R.\ Bulla and S.\ Blawid}
\address{Max-Planck-Institut f\"ur Physik komplexer Systeme,
N\"othnitzer Str. 38, 01187 Dresden, Germany}
\maketitle

\begin{abstract}
We study the extended Hubbard model with both on-site and nearest neighbor
Coulomb repulsion ($U$ and $V$, respectively) in the Dynamical Mean Field
theory. At quarter filling, the model shows a transition to a charge
ordered phase with different sublattice occupancies 
$n_{\rm A}\!\ne\!n_{\rm B}$. The effective mass increases drastically at the 
critical $V$ and a pseudo-gap opens in the single-particle spectral function
for higher values of $V$. The $V_{\rm c}(T)$-curve has a negative slope for 
small temperatures, i.e. the charge ordering transition can be driven by 
increasing the temperature. This is due to the higher spin-entropy of the 
charge ordered phase.

\end{abstract}

}

The possibility of crystallization of electrons due to their long-range
Coulomb repulsion was first proposed by Wigner \cite{Wig38}. He considered
an electron system in a uniform positive background at sufficiently low
densities.
The Wigner lattice is formed when the gain in Coulomb energy due to the
localization of the electrons exceeds the gain in kinetic energy for the 
homogeneous electron distribution.
It is experimentally realized in the two dimensional
electron gas
in a GaAs/AlGaAs heterostructure \cite{Ga}. Due to the
reduced dimensionality, the effect of the Coulomb interaction is enhanced so
that the transition to the ordered state occurs at experimentally accessible
electron densities.

Crystallization of charge carriers (charge ordering) can also be observed
in three dimensional systems, even at very high densities
\cite{Ful97}. Here the kinetic energy of the electrons or holes has to be 
reduced drastically for the charge ordered state to become possible. In 
4f-electron systems it is the small hybridization of the well localized 
4f-orbitals that leads to a reduced kinetic energy. An example is  
$\rm Yb_4As_3$ where a first order charge ordering transition occurs at 
$T_{\rm c}\!\approx\!295$K \cite{Yb4As3,Fulde95}. The carrier concentration in 
$\rm Yb_4As_3$ (approximately one hole per four Yb ions) is considerably 
larger than typical values for a Wigner lattice.
The kinetic energy of the electrons can also be reduced by the
interaction with lattice and spin degrees of freedom. An interplay of
these mechanisms is responsible for the charge order transition
occurring in a variety of rare earth manganites (e.g., in $\rm
La_{1-x}Ca_xMnO_3$ for $x\!\ge\!0.5$ \cite{manganites}). 

In all examples mentioned so far, the charge ordered phase is the ground
state. However, a melting of the charge ordered state on {\it decreasing} the 
temperature (i.e.\ a reentrant transition) has been
found recently in $\rm Pr_{0.65}(Ca_{0.7}Sr_{0.3})_{0.35}MnO_3$
\cite{PrCaSr} and in $\rm LaSr_2Mn_2O_7$ \cite{LaSr}.

In this Letter, we investigate the simplest model which
allows for a charge ordering transition due to the
competition between kinetic and Coulomb energy.
The extended Hubbard model \cite{pen94}
\begin{eqnarray}
H &=& t \sum_{<ij>\sigma} \left(c_{i\sigma}^\dagger c_{j\sigma}
                              + c_{j\sigma}^\dagger c_{i\sigma}\right)  
   -\mu \sum_{i\sigma} c_{i\sigma}^\dagger c_{i\sigma} \nonumber \\
&+& U \sum_i n_{i\uparrow}
          n_{i\downarrow}
     + V \sum_{<ij>} n_i n_j
\label{eq:H1}
\end{eqnarray}
describes fermions on a lattice with an on-site Coulomb repulsion
$U$, a nearest neighbor Coulomb repulsion $V$ and a hopping matrix
element $t$.  The $c_{i\sigma}^\dagger$ ($c_{i\sigma}$) denote 
creation (annihilation) operators for a fermion at site $i$ with spin 
$\sigma$, the $n_i$ are defined as $n_i\!=\!n_{i\uparrow}+n_{i\downarrow}$
where $n_{i\sigma}\!=\!c_{i\sigma}^\dagger c_{i\sigma}$ and $\sum_{<ij>}$
indicates the sum over nearest neighbors. 

In the following we study the extended Hubbard model eq.\ (\ref{eq:H1}) within
the Dynamical Mean Field Theory (DMFT) \cite{Met89,Geo96}, i.e.\ in the limit 
of infinite lattice coordination number $z$. In order to define a nontrivial 
limit as $z\!\to\!\infty$, the parameters $t$ and $V$ are rescaled as 
$t\to t/\sqrt{z}$ and $V\to 2V/z$. This leads to a drastic simplification of 
the self-energy diagrams. The self-energy becomes local and in particular the 
nonlocal Coulomb term contributes only at Hartree level, i.e.\ the $V$-term 
simply acts as a shift of the chemical potential \cite{MH}. Therefore, the 
hamiltonian
\begin{eqnarray}
   H &=& \sum_{i\sigma}\Big( V{\sum_{j}}^\prime \left<n_j\right> -\mu \Big) 
          c_{i\sigma}^\dagger c_{i\sigma} \nonumber \\
&+& t \sum_{<ij>\sigma} \left(c_{i\sigma}^\dagger c_{j\sigma}
             + c_{j\sigma}^\dagger c_{i\sigma}
      \right) + U \sum_i c_{i\uparrow}^\dagger c_{i\uparrow}
          c_{i\downarrow}^\dagger c_{i\downarrow} ,
\label{eq:H2}
\end{eqnarray}
($\sum_{j}^\prime$ indicates the sum over the nearest neighbors of $i$).
leads to the same $z\!\to\!\infty$ limit as eq.\ (\ref{eq:H1}) after
rescaling $t$ and $V$.

As we are interested in charge ordered phases with different occupancies on
the two sublattices A and B of a bipartite lattice,  
we have to generalize the DMFT equations to allow for solutions with 
$n_{\rm A}\!\ne\!n_{\rm B}$. For this, the model (\ref{eq:H2}) is mapped 
self-consistently on two Anderson impurity models (one for sublattice A and 
one for sublattice B). In the Bethe lattice case, where the density of states 
is given by $D(\epsilon)\!=\!\frac{2}{\pi W^2} \sqrt{W^{2}-\epsilon^{2}}$ 
(we set $W\!=\!1$ as the unit for the energy scale), the self-consistency 
equations simplify to the form:
\begin{eqnarray}
  \epsilon_{d}^{\rm A}+
  \Delta_{\rm A}(i\omega_n) & = &
  2Vn_{\rm B}-\mu+\frac{1}{4}G_{\rm B}(i\omega_{n}) ,
  \label{eq:S1} \\[3mm]
  \epsilon_{d}^{\rm B}+
  \Delta_{\rm B}(i\omega_n) & = &
  2Vn_{\rm A}-\mu+\frac{1}{4}G_{\rm A}(i\omega_{n}),
\label{eq:S2}
\end{eqnarray}
($G_{\rm A/B}(i\omega_{n})$ are the Green functions for the A/B 
sublattice and $\Delta_{\rm A/B}(i\omega_{n})$ are the hybridization
functions between impurity and effective conduction band;
$n_{\rm A/B}$ denote the sublattice occupancies and $\epsilon_{d}^{\rm A/B}$ 
are the on-site energies of the effective impurities).

The remaining problem is the 
iterative solution (i.e.\ the calculation of the Greens
functions) of an effective single impurity Anderson model. We use
three different methods, an Exact Diagonalization technique (ED)
for finite temperatures,
the Non-Crossing Approximation (NCA) and the Numerical Renormalization 
Group method (NRG). 
The ED method diagonalizes an impurity model with a finite number 
$N$ of conduction band orbitals. Within the self-consistency procedure we 
define the mapping of the full Green functions $G_{\rm A/B}(i\omega_{n})$ to 
the hybridization functions $\Delta_{\rm B/A}(i\omega_{n})$ by expanding both 
sides of (\ref{eq:S1}),(\ref{eq:S2}) in powers of $(i\omega_{n})^{-1}$ and 
match coefficients up to order $(i\omega_{n})^{-2N}$. This approximation is 
similar in spirit to the projection method based on the continued fraction 
representation used for $T\!=\!0$ \cite{Geo96}. The NRG is applied here for 
the first time to a particle-hole asymmetric problem within the DMFT. The 
method is an extension of earlier work on the (not extended) symmetric
Hubbard model \cite{BHP}. Details of the NCA approach are summarized
in \cite{Stefan}.

\begin{figure}[t]
\epsfxsize=3.0in
\epsffile{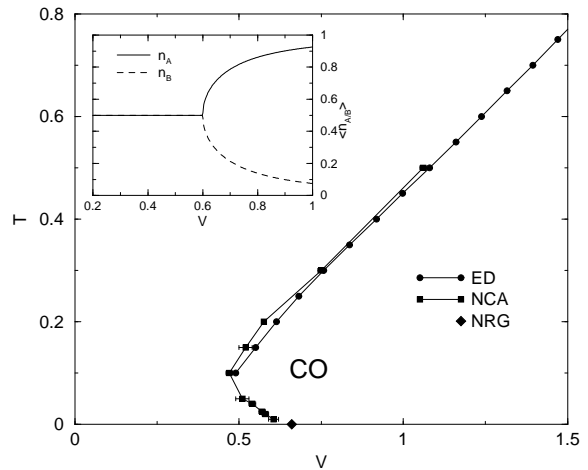}
\caption[]{Phase diagram for the extended Hubbard model ($U\!=\!2$, quarter
           filling). The various symbols show the results for the phase
           boundary between the homogeneous and the charge ordered phase (CO)
           from ED (circles), NCA (squares) and NRG calculations
           (diamond). The inset shows the ED result for the $V$-dependence of 
           the lattice occupancies for $U\!=\!2$ and $T\!=\!0.2$.}
\label{fig:phased}
\end{figure}

Fig. \ref{fig:phased} shows the ($T$-$V)$-phase diagram for
$U\!=\!2$ and quarter filling.
The results of the different methods agree remarkably
well in their corresponding range of applicability. 
For high temperatures ($T\!>\!0.4$), the ED results for $N\!=\!5$ 
shown in Fig.\ \ref{fig:phased} can already be obtained from an $N\!=\!3$ 
calculation (within numerical accuracy). The ED method cannot be used for 
very low temperatures since only a small number of orbitals is taken into 
account. The NCA is applicable down to much lower temperatures and we find 
that the slope of the $V_{\rm c}(T)$-curve changes its sign at 
$T\!\approx\!0.1$. The NCA encounters problems in the very low temperature 
limit. Nevertheless, the extrapolation of the $V_{\rm c}(T)$-curve to 
$T\!=\!0$ agrees well with the critical $V$ obtained from the NRG method 
$V_{\rm c}(T=0)\!\approx\!0.66$.

The reentrant behavior in the interval $0.47\!<\!V\!<\!0.66$ is due to the
higher spin entropy of the charge ordered state. As we have neglected the
possibility of an additional magnetic ordering in the charge ordered
state, the contribution of the spin degrees of freedom to the entropy is 
approximately $\ln(2)$ per occupied site. The entropy of the homogeneous 
phase, which is a Fermi liquid, grows linearly with $T$.
On further increasing the temperature, the increasing charge entropy of the
homogeneous phase dominates and the system shows the usual melting behavior.

The inset of Fig.\ \ref{fig:phased} shows the ED-result for the
$V$-dependence of the sublattice occupancies $n_{\rm A}$ and $n_{\rm B}$
($T\!=\! 0.2$, $U\!=\!2$).
The transition is clearly continuous, in contrast to the result for 
$T\!=\!0$ where the NRG gives a first order phase transition with a jump
in the order parameter $\delta n \!=\!n_{\rm A}-n_{\rm B}$ from
0 to $\delta n \!\approx\! 0.7$. Unfortunately, it is not possible to
clarify numerically how the first order transition at $T\!=\!0$
evolves from the continuous transition at finite $T$. 
There are numerical indications that $\delta n$ increases
more rapidly at the transition when $T$ is reduced.
However, the convergence of the iterative procedure is extremely slow 
near the phase boundary. 

\begin{figure}[tbp]
\epsfxsize=3.0in
\epsffile{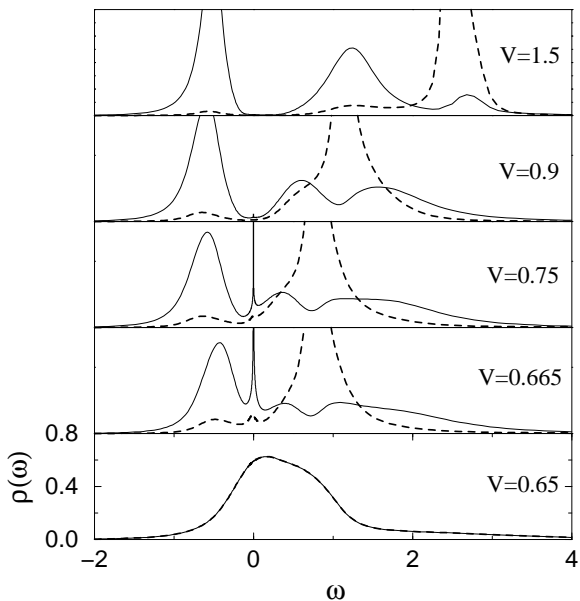}
\caption[]{NRG-results for the A and B spectral functions $\rho(\omega)$
           (solid and dashed lines, respectively) for
           $U\!=\!2$, $T\!=\!0$ and
           different values of $V$. Below
           the critical $V_c\!\approx\!0.66$, the spectral functions
            of both sublattices are equal.}
\label{fig:NRG}
\end{figure}

The NRG results for the A and B spectral functions for $T\!=\!0$
are shown in Fig.\ \ref{fig:NRG}. Below $V\!=\!V_{\rm c}$, the A and B 
spectral functions are equal and independent of $V$ (the 
extended Hubbard model in the DMFT reduces to the ordinary
Hubbard model as long as the homogeneous phase is considered; the solution
shown in Fig.\ \ref{fig:NRG} for $V\!=\!0.65$ is therefore also 
the solution for the Hubbard model with $V\!=\!0$). The discontinuous 
transition to a finite $\delta n$ at $V_{\rm c}$ is reflected in a 
redistribution of spectral weight in the spectral functions above $V_{\rm c}$. 

For $V\!=\!0.665$ the system is still a Fermi liquid with a strongly enhanced 
effective mass ($m^\ast/m\!\approx\!120$) and a very narrow quasiparticle 
resonance at the Fermi level. The effective mass increases further 
on increasing $V$ and an extreme narrow quasiparticle resonance is still 
present for $V\!=\!1.5$ (the width of this resonance is below the energy 
resolution used in Fig.\ \ref{fig:NRG} so that the peak is not visible).
Although the system is still a Fermi liquid on an extremely small energy scale,
a pseudo-gap in the A and B spectral functions gradually
develops at $V\!\approx\!0.9$. The spectral weight within the pseudo-gap 
decreases approximately as $1/V^2$ for large $V$. 

\begin{figure}[t]
\epsfxsize=3.0in
\epsffile{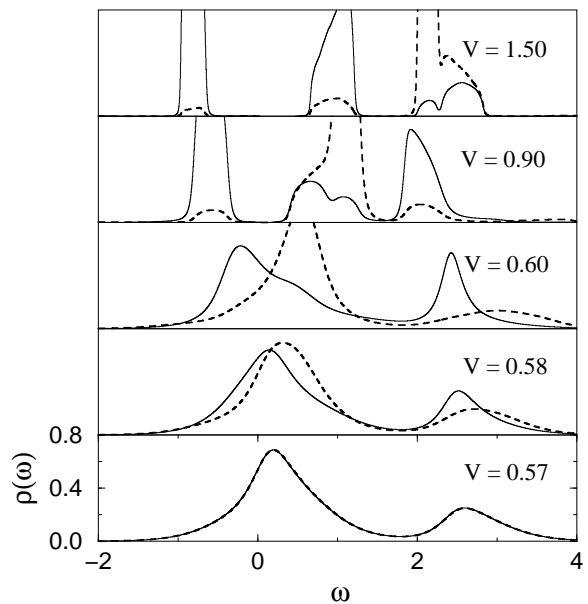}
\caption[]{NCA-results for the
A and B spectral functions $\rho(\omega)$ (solid and dashed lines,
respectively) $U\!=\!2$, $T\!=\!0.2$ and
different values of $V$. For $V\!\geq\!0.58$ charge order
sets in and a pseudo-gap develops at the Fermi level.}
\label{fig:NCA}
\end{figure}

The NCA-result for the spectral function for $U\!=\!2$, $T\!=\!0.2$ and 
various values of $V$ is shown in Fig.\ \ref{fig:NCA}. The very narrow 
quasiparticle peak seen at $T\!=\!0$ is absent because the temperature 
exceeds the low energy scale associated with the Fermi liquid behavior. 
In contrast to the $T\!=\!0$-case, the transition is continuous which is 
reflected in the continuous transfer of spectral weight at the transition. 
A comparison of the NRG and NCA results for $V\!=\!0.9$ and $V\!=\!1.5$ shows 
a good agreement of the main features of the spectral functions. However,
the resolution of the high energy features is limited in the NRG since the
spectra are obtained by broadening a discrete set of $\delta$-peaks.

In the homogeneous phase, the NCA spectral function shows the splitting in 
upper and lower Hubbard band. The weight of the upper Hubbard band in the 
B-spectral function is considerably larger compared to the NRG result for 
$T\!=\!0$. The weight decreases on increasing $V$ in the charge ordered phase 
because the occupancy of the B-sites is suppressed. In addition, the bands 
in the spectral function narrow due to the reduced hopping of the electrons 
in the charge ordered phase. In the limit of large $V$, perfect charge order 
evolves ($n_{\rm A}\!\to\!1$ and $n_{\rm B}\!\to\!0$). Therefore, in this
limit the total spectral function of the extended Hubbard model has a 
simple three-peak structure with peaks at $E_1\!=\!-\mu$, $E_2\!=\!U-\mu$ 
and $E_3\!=\!2V-\mu$. The peak at $E_1$ corresponds to the process of taking 
out one electron from a (singly occupied) A-site, the peak at $E_2$ is due 
to the addition of an electron to an A-site and the peak at $E_3$ comes from 
the addition of an electron to a B-site.

\begin{figure}[t]
\epsfxsize=3.0in
\epsffile{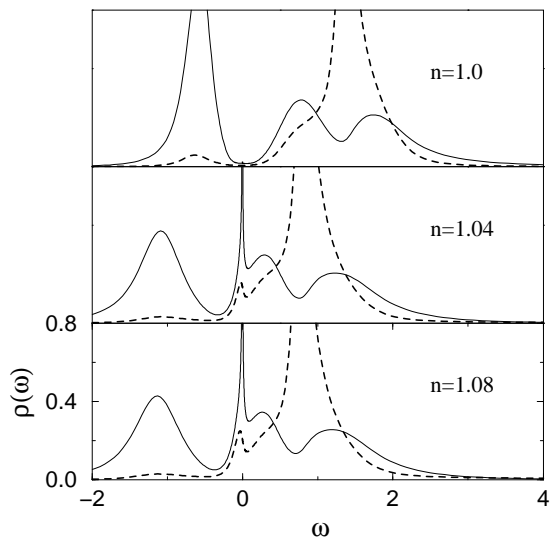}
\caption[]{NRG-results for the A and B spectral functions $\rho(\omega)$
          (solid and dashed lines, respectively) for $V\!=\!1.0$,
           $T\!=\!0$ and different occupancies $n$. A transfer of weight
           from the lower Hubbard band (of both A and B spectral functions)
           to the Fermi level is clearly visible.}
\label{fig:dope}
\end{figure}

It is tempting to explain the opening of the charge-order gap by the
change of the unit cell alone, reflecting the broken AB-symmetry of
the lattice. Although the AB-symmetry is of course broken at the
charge order transition, our results for the spectral function cannot
be understood within a one-particle picture. First of all, the
charge-order gap visible in the spectral function is not a real hard
gap but only a pseudo-gap (i.e.\ an energy-regime of strongly reduced
spectral weight) even at zero temperature. From an experimental point
of view this difference is irrelevant. However, a pseudo-gap cannot be
obtained by a simple band splitting due to a reduced symmetry. To
demonstrate the many-particle character of the bands, we have doped
the extended Hubbard model away from quarter-filling 
(see Fig.\ \ref{fig:dope}). Note that within our approach the charge ordered
state on the A/B-sublattice exists also for $n\!\neq\!1$.
A rapid weight transfer between the high- and the 
low-energy scale is observed, which is a characteristic feature 
of strongly correlated electron systems. A well examined example is the usual 
half filled Hubbard model, where the weight of the lower Hubbard band 
decreases faster than the doping $x$ \cite{Harris67,Eskes94b}. In the extended
Hubbard model we find the same behavior. 

In conclusion, we have found that the charge order transition in the infinite
dimensional extended Hubbard model shows a number of very interesting
features. For low temperatures, the system is a Fermi liquid for {\it
both} $V\!<\!V_c$ and $V\!>\!V_c$. Nevertheless, the effective mass undergoes
a rapid change at the transition. The corresponding very narrow
quasiparticle peak at the Fermi energy will hardly be seen in
photoemission experiments of charge ordered materials. However, the huge
mass enhancement at the transition has drastic consequences on the
transport properties, e.g.\ on the resistivity. This is the case in 
$\rm Yb_4As_3$, where the charge order transition is accompanied by a jump in 
the resistivity.

In our model calculations, we find a real first order transition only
for $T\!=\!0$, although the transition remains sharp also for small 
$T\!\neq\!0$. The occurrence of a first order transition at {\it finite}
temperatures in the above mentioned examples indicates that lattice degrees 
of freedom are involved. Indeed, the charge order transition in 
$\rm Yb_4As_3$ is accompanied by a structural phase transition.

For a particular range of the nearest neighbor Coulomb repulsion $V$ the
extended Hubbard model shows reentrant behavior, i.e.\ a transition
from the charge ordered to the homogeneous phase with {\it decreasing}
temperature. This behavior is experimentally seen in $\rm
Pr_{0.65}(Ca_{0.7}Sr_{0.3})_{0.35}MnO_3$ \cite{PrCaSr} and in $\rm
LaSr_2Mn_2O_7$ \cite{LaSr}. Moreover, the bandwidth in $\rm
Pr_{0.65}(Ca_{1-y}Sr_y)_{0.35}MnO_3$ can be varied by doping with Sr
or by applying a magnetic field. For sufficiently small kinetic energy
of the electrons only one charge order transition occurs. Increasing
the kinetic energy, a reentrant regime is found until for high dopings
or magnetic fields the charge order is totally suppressed. Although
the localization mechanisms in the rare-earth manganites are more
complex than the one induced by a nearest neighbor Coulomb repulsion,
we find the same behavior in our model as function of $V/W$. We take
this as an indication that the occurrence of a reentrant charge order
transition does not depend on details of the localization mechanism.

Finally, in photoemission experiments of charge ordered materials the
many-particle character of the system should be seen in a
weight-transfer upon doping with additional charges. 

The authors would like to thank P.\ van Dongen and
P.\ Thalmeier for helpful discussions.

\end{document}